\documentstyle[sprocl]{article}

\begin{document}

\title{Particle Mixing and Charge Asymmetric $\Lambda N$ Forces}

\author{S.\ A.\ Coon, H.\ K.\ Han}
	
\address{Physics Department,
        New Mexico State University, Las Cruces, NM  88003, USA}

\author{	 J.\ Carlson,
	B.\ F.\ Gibson}
	
\address{	Theoretical Division,
	Los Alamos National Laboratory, Los Alamos, NM 87545, USA}

\maketitle\abstracts{We calculate the contributions of a particular set
of charge asymmetric $\Lambda N$ interactions to the difference of the 
separation energies of $^4_{\Lambda}$He and $^4_{\Lambda}$H. We use
perturbation theory with four-body variational Monte Carlo wave
functions calculated from a Hamiltonian with two- and three-hadron
forces. We compare with the data and with an earlier calculation made
by one of us which employed  a two-body wave function of the
$\Lambda$-nucleus type.}

\section{Introduction}

Dalitz and Von Hippel (DvH)~\cite{DvH} applied  the Coleman-Glashow
analysis of electromagnetic mass splittings~\cite{CG} to 
 the charge asymmetry (CA) of the $\Lambda N$ interaction,
long before the acceptance of $\rho^0 \omega$ mixing as a major
contribution to  nuclear charge asymmetry~\cite{CB,Miller} .  Recently
the dominant role of meson mixing in the CA $NN$ interaction has been
questioned~\cite{Thomas}, but this issue is not yet finally
settled~\cite{Miller,CMA}.   We reexamine particle mixing
as a mechanism for charge asymmetry in the $\Lambda N$ interaction. 
 In this contribution, we adopt  the single meson exchange charge asymmetric
potentials~\cite{CM} based upon the Coleman-Glashow $SU(2)$ breaking
tadpole  mechanism~\cite{CG}.  They include the DvH  potential
of one-pion exchange range allowed by $\Sigma^0-\Lambda$ mixing, the
  concomitant rho-exchange potential, and
 $\Lambda N$  potentials from $\rho^0 \omega \phi$ and
$\pi^0 \eta \eta'$ mixing.  

\section{Charge asymmetry in $^4_{\Lambda}$He and $^4_{\Lambda}$H}

The strongest experimental evidence for a violation of charge symmetry lies in
the $\Lambda$-separation  energies $B_{\Lambda}$ of the four-body
mirror hypernuclei   $^4_{\Lambda}$He and $^4_{\Lambda}$H.  The separation
energy of the $\Lambda$ in a hypernucleus is defined as  
\begin{equation}
  B_{\Lambda} = M(^{A-1} Z) + M_{\Lambda} - M(^A_{\Lambda} Z)\,\, ,
  \label{eq:sepeng}
\end{equation}
where $M(^{A-1} Z)$, $M_{\Lambda}$, and $M(^A_{\Lambda} Z)$ are the
masses of the core nucleus, hyperon, and hypernucleus, respectively. 
We write $B_{\Lambda}$ as the difference of the total energies of the
hypernucleus and core nucleus ($ E(^3{\rm H}) \approx -8.48$ MeV, for
example)
\begin{equation}
B_{\Lambda} = - [E(^A_{\Lambda} Z) - E(^{A-1} Z) ]\,\,.
\label{eq:bediff}
\end{equation}
From Eq. (\ref{eq:bediff}) it is clear that the separation
energies of $\Lambda$'s in mirror hypernuclei will be equal if the 
$\Lambda N$ force is charge symmetric and if the $\Lambda$-particle does
not distort the core nucleus.  The presence of a bound $\Lambda$-particle
tends to compress the core nucleus, thus increasing the Coulomb repulsion
 of $^4_{\Lambda}$He, which in turn decreases the value of 
$B_{\Lambda} (^4_{\Lambda} {\rm He})$. Therefore, the Coulomb effect from the 
distortion of the core is negative, and can be calculated to first order
as $\Delta B^C_{\Lambda} = -\Delta E^C_{\Lambda}$ where $\Delta E^C_{\Lambda}
 = E_C( ^4_{\Lambda}
{\rm He}) - E_C(^3 {\rm He})$.  It has 
been estimated to be about -20 keV, by analogy with the  core
compression suggested by empirical evidence on the Coulomb energies of
$^3$He and (the compressed by one more neutron) $^4$He~\cite{Coulomb}. 
 Experimentally,
\begin{equation}
  \Delta B_{\Lambda}^{{\rm exp}} =  B_{\Lambda} (^4_{\Lambda} {\rm He}) -
    B_{\Lambda} (^4_{\Lambda} {\rm H}) = +350 \pm 50 \,\,\,{\rm keV},
    \label{eq:gsdeltal}
\end{equation}   
from the ground state separation energies 2.39 $\pm$ 0.03 MeV and 
2.04 $\pm$ 0.04 MeV~\cite{gs}.  
The theoretical $\Delta B_{\Lambda}$ to be attributed to charge
asymmetry 
($\Delta B_{\Lambda} = \Delta B_{\Lambda}^{{\rm exp}} - \Delta B^C_{\Lambda}$)
is then slightly greater than $\Delta B_{\Lambda}^{{\rm exp}}$.
A second measure of charge asymmetry is provided by
observations~\cite{exstates} of the M1 $\gamma$-rays from the first
excited states:
\begin{equation}
  \Delta B_{\Lambda}^{*{\rm exp}} \approx  +230 \,\,\,{\rm keV},
    \label{eq:esdeltal}
\end{equation} 
The excited states have $J^\pi = 1^+$ instead of the $J^\pi = 0^+$
assignment of the ground states, and  the different four-body states
contain significantly different mixtures of the spin-singlet and
spin-triplet  $\Lambda N$ interaction~\cite{Carlson}.  In a simple
model, the excited state is  characterized by a spin-flip of the
$\Lambda$ from its ground state configuration.  In any model, the two
measures of charge asymmetry in Eqs. (\ref{eq:gsdeltal}) and
(\ref{eq:esdeltal}) (increased slightly by the calculated  $\Delta
B^C_{\Lambda}$) can be compared with the predictions of theoretical charge
asymmetric $\Lambda N$ potentials.

\section{Variational Monte-Carlo calculations}

Early perturbative estimates~\cite{DvH,CM} of the effects of CA relied
on a two-body wave function obtained from a folding model of the
$\Lambda$ nucleus interaction. Such a two-body calculation has been
shown to  differ from an exact four-body calculation~\cite{Gibson}.
Therefore we employ  a variational Monte Carlo method developed for
accurate numerical calculations of light nuclei~\cite{CPW}.  The
two-hadron potentials have been adapted to this calculational scheme
(extended~\cite{Carlson} to the four hadron bound state in which one
hadron could be a $\Lambda$). The ``Urbana-type   potentials", suited
to this variational approach, take the form of  a sum of operators
multiplied by functions of the interparticle distance. We use the
original (charge symmetric) Reid soft core nucleon-nucleon potential in
the form of the Urbana-type Reid $v_8$ potential~\cite{reidv8}. The
charge symmetric $\Lambda N$ potential is  a spin dependent potential
of the Urbana type~\cite{csvlam,Bodmer}.  To achieve a good description
of the core nuclei $^3$H and $^3$He, we employ two-parameter
three-nucleon forces of a similar Urbana type, introduced in Ref. 12.
The Reid $v_8$ is a simplified (the sum of operators is truncated from
a possible 18 to  8 operators) $NN$ force model which is equivalent to
the original Reid soft core nucleon-nucleon potential in the lower
partial waves and can produce the dominant correlations in s-shell
nuclei.  In a similar fashion,  the Urbana $NNN$ potentials feature a
short range repulsion of purely central character plus  only those
operators corresponding to two-pion exchange which  produce significant
contributions to the  binding energies of s-shell nuclei. It has been
demonstrated that one can obtain a quantitatively satisfactory
description  of the core nuclei $^3$H and $^3$He with such a model. 
Even though the set of $NNN$ operators has been truncated, a need to
include the additional operators of the two-pion-exchange NNN
interaction~\cite{CPW,Friar} or of other theoretical $NNN$ potentials
has not been compelling. Therefore, the combination of 
two and three hadron potentials utilized in this study provide a
convenient means of obtaining accurate four-hadron wave functions for
the purpose of perturbative estimates of the effects of charge
asymmetric  $\Lambda N$ potentials in the four-body mirror
hypernuclei   $^4_{\Lambda}$He and $^4_{\Lambda}$H.  To this end, we
also include a $\Lambda NN$ interaction, again  of the Urbana
type~\cite{lnntbf}.

 Variational calculations with the above charge symmetric potentials
give a satisfactory description of the non-strange core nuclei. In Table
1, we
average,  over $^4_{\Lambda}$He and $^4_{\Lambda}$H hypernuclei, the
ground state $\langle^4_{\Lambda}\rangle$ and  first excited state
$\langle ^4_{\Lambda}(*)\rangle$ energies to suggest that the
variational wave functions should be adequate for perturbation theory.
\begin{table}[h]
\caption{Calculated total energies in MeV with charge symmetric potentials.
\label{tab:props}}
\vspace{0.2cm}
\begin{center}
\footnotesize
\begin{tabular}{|c r r|}
\hline
{} & E(exp)  & E(model) \\
 $^3$H &  -8.48  & -8.27 \\
 $^3$He &  -7.72 & -7.63 \\
 $^4$He &  -28.30 & -28.39 \\
  $\langle^4_{\Lambda}\rangle$ & -10.32 & -9.77 \\
 $\langle ^4_{\Lambda}(*)\rangle$ & -9.20 & -8.58 \\ 
\hline
\end{tabular}
\end{center}
\end{table}
 \noindent Note the reasonable description of the
$\approx 1$ MeV spacing between the (averaged) ground state and the
spin-flip excited state given by our chosen interactions.
 These energies were obtained with the Urbana IX $NNN$
potential, the latest in a series of phenomenological $NNN$
interactions~\cite{nnntbf}.

The predicted Coulomb energy of $^3$He is 644 keV, exactly what would
be expected from the measured charge form factors of $^3$He and
$^3$H~\cite{Coulomb,BCS}, indicating that our charge symmetric Hamiltonian
provides a good description of the charge
radii of the two core nuclei.  The Coulomb contribution, $ \Delta B
^C_{\Lambda}$,
to $\Delta B_{\Lambda}^{{\rm exp}}$ of Eq. (\ref{eq:gsdeltal}) is -42 keV, a
little more than the estimate of Ref. [8], but consistent with an
earlier variational study~\cite{Bodmer}.  From our calculations, the
Coulomb contribution to the excited state  $\Delta B_{\Lambda}^{*{\rm
exp}}$
of Eq. (\ref{eq:esdeltal}) is -28 keV.  Both estimates are obtained by
subtracting two numbers, each with its Monte Carlo uncertainties, so
they should be taken as indicative rather than definitive. Still, it is
clear that the $\Delta B_{\Lambda}$ to be attributed to charge
asymmetry ($\Delta B_{\Lambda} = \Delta B_{\Lambda}^{{\rm exp}} - 
\Delta B^C_{\Lambda}$)  is positive for both the ground and excited
states and  the Coulomb corrections to the experimental values  are
small.

\section{Particle mixing and $\Delta B_{\Lambda}$ }

In this contribution, we wish to demonstrate the change in  $\Delta
B_{\Lambda}$ (calculated as a first order perturbation of the charge
asymmetric potential) as one upgrades a plausible ${\Lambda}$-plus-core
nucleus model~\cite{DvH,CM} to a four-body model obtained variationally
from plausible potentials.  The variational parameters of the
${\Lambda}$-plus-core-nucleus model (labelled 2-body in Table 2 below)
were fit to the (averaged) charge radii of the core nuclei and to the
total energy of $_\Lambda^4$H (for details see Eqs. (46-48) of Ref. 7) 
Thus the long range aspects of the two-body wave function were
determined about as well as those of the variational Monte Carlo wave
function of the previous section (labelled 4-body in Table 2 below). 
The short range correlations of the two-body wave function were
simulated by a correlation  function which was zero inside a
``hard-core" radius of 0.4 fm and unity elsewhere.  In the present
4-body wave function the two- and three-body correlations are determined
variationally by the short range behaviour of the two- and three-body
potentials.

We fix the mixing constants and coupling constants of the charge
asymmetric potentials of Ref. 7 at the values determined in 1979,  even though our
knowledge of these numbers has evolved over the last twenty years.  
These published potentials  have central, spin-spin and tensor
operators, but the 2-body wave function used there had only $S$-wave
components.  So we evaluate only the central and spin-spin operators
with the present  4-body wave function.  The results are compared in
Table 2.

\begin{table}[h]
\caption{Contributions in keV to $\Delta B_{\Lambda}$ from the potentials
due to particle mixing \label{tab:contributions}}
\vspace{0.2cm}
\begin{center}
\footnotesize
\begin{tabular}{|l r r r r|}
\hline
   & \multicolumn{2}{c} {$J^\pi = 0^+$} 
& \multicolumn{2}{c|}{$J^\pi = 1^+$}   \\
\hline
{  } &  4-body  &  2-body  &  4-body  & 2-body \\
  &  &   &   &  \\
$\Delta B_{\pi - {\rm exchange}}^{\Sigma \Lambda}$ & 45 & 59 & -11  &
-20 \\
 $\Delta B_{\rho - {\rm exchange}}^{\Sigma \Lambda}$ & 75 & 112 & -11  &
 -52 \\
 $\Delta B^{\pi \eta}$ & -7 & -18 & 1  & 6 \\
 $\Delta B^{\rho \omega}$ & -14 & 22 &  -10 & 23 \\
 $\Delta B^{\rho \phi} $ & -8  & -1 & -2  & 7 \\
  &  &   &   &  \\
   Total ($\Delta B_{\Lambda}) $ & 91 & 174 & -33  & -36 \\
 \hline
\end{tabular}
\end{center}
\end{table}

We note that the simple 2-body ${\Lambda}$-plus-core-nucleus model might
be  adequate for evaluation of the long range charge asymmetric force
allowed by the $\Sigma^0 \Lambda$ transition,  but the potentials of
shorter range are badly served by this model.  The effect of the
previously dominant single $\rho$-exchange potential is especially
lowered by the 4-body evaluation.

Now we turn to a confrontation of Table 2 with the experimental values
of Eqs. (\ref{eq:gsdeltal}) and (\ref{eq:esdeltal}).  These particle
mixing potentials do not explain the large charge asymmetric effect in
the s-shell hypernuclei. In particular they appear to have a
spin dependence quite different from experiment.

 One effect left out of this calculation is a possible
contribution~\cite{Dalitz} of small $D$-components of the hypernuclear
wave function which have a strong coupling to the dominant
$S$-components through  the tensor operator of the charge asymmetric 
$V_{\pi - {\rm exchange}}^{\Sigma \Lambda}$,  $V_{\rho - {\rm
exchange}}^{\Sigma \Lambda}$,and $V^{\pi \eta}$.  Our  charge symmetric
$\Lambda N$ was assumed to be central for this first four-body
calculation and therefore cannot create such $D$-components.  We hope
to report,in a future publication, calculations with a more general
charge symmetric  $\Lambda N$ interaction and an update of the
Coleman-Glashow mixing and of the coupling constants of the charge
asymmetric $\Lambda N$ interaction.

For the present, we are left with a larger charge asymmetry to explain
in the $\Lambda N$  interaction than in the $N N$ interaction.  The
model charge asymmetry based on particle mixing, which works extremely
well in the $N N$ interaction,  appears to be inadequate to explain all
the charge asymmetry in hypernuclei.  These facts once again emphasize
that hypernuclear physics is not simply nuclear physics with the word
``hyper" added.  There are new surprises and new mysteries in strange
nuclear systems.

\section*{Acknowledgments}
The work of JAC
 and BFG was supported by  the U. S. Department of Energy. The work
 of SAC was supported in part by NSF grant No. PHY-9722122.  This
 work was done while HKH was a visitor at Los Alamos National
 Laboratory and NMSU, supported in part by the U. S. Department of
 Energy, National Science Foundation, and the Korean Scientific and
 Engineering Foundation.

\end{document}